\documentclass[
    reprint,
    floatfix,
    amsmath,
    amssymb,
    aps,
    pra,
 ]{revtex4-2}

\usepackage[english]{babel}

\usepackage{graphicx}
\usepackage{dcolumn}
\usepackage{bm}
\usepackage{xcolor}

\usepackage[
    colorinlistoftodos,
    color=green!40,
    prependcaption
]{todonotes}
\usepackage[hidelinks]{hyperref}

\newif\ifshowchanges
\showchangesfalse 

\newcommand{\rev}[1]{%
    \ifshowchanges
        \textcolor{magenta}{#1}%
    \else
        #1%
    \fi
}

\begin{document}
\raggedbottom
\preprint{APS/123-QED}
\title{Single-Mode Squeezing Using Degenerate Four-Wave Mixing in a Dual-Pump TWPA}

\author{Anna Levochkina}
\email{Corresponding author: anna.levochkina@aalto.fi}
\author{Kirill Petrovnin}
\author{Iiro Harju}
\author{Jiaming Wang}
\author{Gheorghe Sorin Paraoanu}


\affiliation{Department of Applied Physics, Aalto University, FI-00076 Aalto, Finland}

\date{\today}

\begin{abstract}
Single-mode squeezing of the electromagnetic vacuum is a key resource for mode-selective noise reduction in quantum measurements. At microwave frequencies, single-mode squeezing is typically obtained by using Josephson and Traveling-Wave Parametric Amplifiers designed for three-wave mixing operation.
\rev{In single-pump degenerate four-wave mixing, the coincidence of pump, signal, and idler frequencies complicates separation of the squeezed field from the strong pump. A dual-pump configuration enables degenerate operation while keeping both pumps spectrally separated from the squeezed mode.}
\rev{Here, we demonstrate single-mode squeezing in a flux-tunable, dual-pump Traveling-Wave Parametric Amplifier, employing reversed-Kerr phase matching.}
\rev{By independently calibrating the single-pump gains, we find that the strongest squeezing occurs along approximately balanced-gain trajectories, while the optimal squeezing does not necessarily correspond to the highest dual-pump gain.}
\rev{At flux biases $\Phi_\mathrm{ext}/\Phi_0 = 0.0$ and $0.5$, the minimum quadrature noise levels reach $-2.48 \pm 0.18~\mathrm{dB}$ and $-1.62 \pm 0.20~\mathrm{dB}$ relative to vacuum, respectively.}
\rev{These results establish the proposed architecture as a platform for single-mode squeezing in microwave continuous-variable quantum technologies.}

\end{abstract}


\maketitle


\section{Introduction} 
\label{sec:introduction}

Squeezing of the electromagnetic vacuum is a quantum phenomenon in which vacuum fluctuations of the electromagnetic field are redistributed between conjugate quadratures, leading to reduced uncertainty in one quadrature below the standard vacuum limit and a corresponding increase in the other, in agreement with the Heisenberg uncertainty principle \cite{Walls_1983, Walls, Caves_1981}. It plays a central role in quantum metrology and continuous-variable quantum technologies by enabling reduced quantum noise and enhanced measurement sensitivity \cite{Wallraff, Aoki, Jensen}.

In parametric systems, squeezing is induced by a strong pump field ($f_\text{p}$) that mediates energy exchange between the signal ($f_\text{s}$) and idler ($f_\text{i}$) modes. The relation between $f_\text{p}$, $f_\text{s}$, and $f_\text{i}$ depends on the underlying nonlinear mixing process \cite{Esposito_perspectives}. In three-wave mixing (3WM), a single pump photon generates one signal and one idler photon, satisfying $f_\text{p} = f_\text{s} + f_\text{i}$. In four-wave mixing (4WM), two pump photons combine to produce the signal and idler, such that $2f_\text{p} = f_\text{s} + f_\text{i}$. These nonlinear processes produce different forms of squeezing depending on the spatial relation of the signal and idler modes. 

\begin{figure*}[t]
    \centering
    \includegraphics[width=0.87\textwidth]{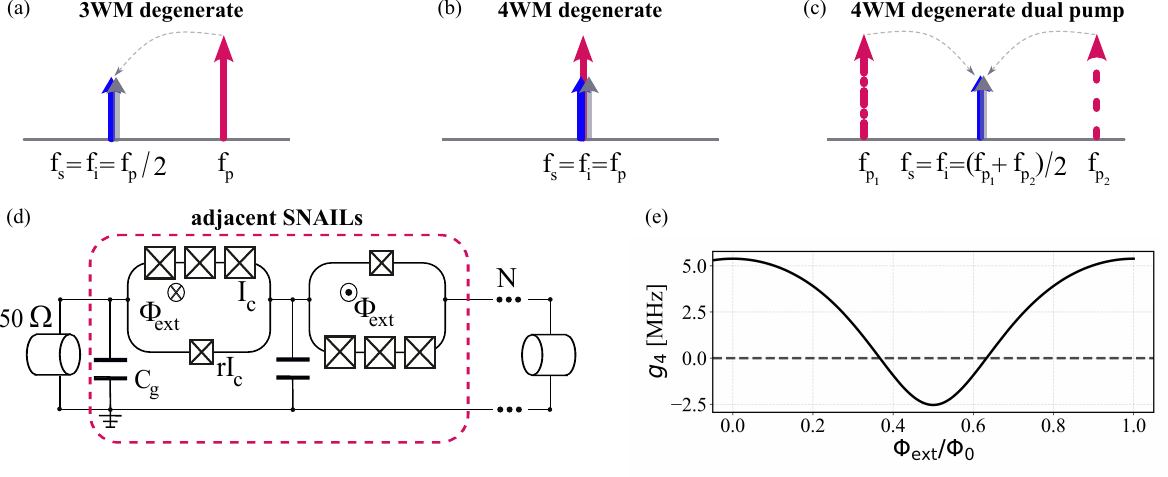}
    \caption{
    \textbf{Schematic representation of the parametric mixing processes and device characteristics.}
    (a) Three-wave mixing (3WM),
    (b) four-wave mixing (4WM),
    (c) 4WM in a dual-pump scheme.
    Here, $f_\text{p}$, $f_\text{s}$, and $f_\text{i}$ denote the pump, signal, and idler frequencies, respectively.
    (d) Reversed-Kerr TWPA \cite{Ranadive} comprising 350 unit-cell
    pairs (700 SNAILs), with two SNAILs shown in the inset.
    (e) \rev{Bare} quartic coefficient $g_4$ versus normalized
    external flux $\Phi_\mathrm{ext}/\Phi_0$.
    }
    \label{fig:1}
\end{figure*}

When the signal and idler coincide in frequency, the process produces single-mode squeezing \cite{Semmler}. In contrast, when the signal and the idler occupy distinct modes, their correlations cause two-mode squeezing, which can generate entanglement \cite{Eichler_11}. \rev{Two-mode squeezing has been demonstrated in both 3WM and 4WM regimes in superconducting microwave platforms, including Josephson Parametric Amplifiers (JPAs) \cite{Malnou, Krantz, Shaw26} and Traveling-Wave Parametric Amplifiers (TWPAs) \cite{Grimsmo, Perelshtein, Esposito}.}
\rev{Single-mode squeezing is commonly implemented in the 3WM regime, where the pump is spectrally separated from the squeezed field and can therefore be efficiently filtered (Fig.~\ref{fig:1}(a)). Unlike in 3WM, single-pump degenerate 4WM produces squeezing at the pump frequency ($f_p = f_s = f_i$), preventing the strong pump from being spectrally separated from the squeezed field (Fig.~\ref{fig:1}(b)). This limitation motivates alternative pumping schemes for realizing single-mode squeezing in 4WM devices, particularly since many broadband parametric amplifiers operate in this regime \cite{Macklin_15, Ranadive, Peng_22}.}

One such approach was recently demonstrated in the work of Ref. \cite{Brein}, where degenerate 4WM was realized using a dual-pump scheme (Fig.~\ref{fig:1}(c)). The scheme employs two appropriately detuned pump tones  $f_{\text{p}_1}$ and $f_{\text{p}_2}$ enable phase-sensitive amplification while suppressing pump leakage at the signal frequency $f_\text{s} = \frac{f_{\text{p}_1} + f_{\text{p}_2}}{2}$, resulting in high-quality single-mode squeezing.  The primary challenge in this method is to achieve proper phase-matching, which in that work is based on dispersive engineering \cite{obrien2014resonant, malnou2025twpa, white2015traveling}. While this approach delivers excellent performance, it imposes constraints on the choice of pump frequencies and operational flexibility. This raises the question of whether single-mode squeezing can be realized in TWPAs that employ alternative phase-matching mechanisms. 

Here, we investigate this possibility using a TWPA in which phase-matching is achieved via a sign reversal of the Kerr nonlinearity \cite{Ranadive_PhD}. 
The device under test is the Carthago amplifier \cite{SinetWaves}, a flux-tunable TWPA implemented as a transmission line composed of Superconducting Nonlinear Asymmetric Inductive eLements (SNAILs) \cite{Frattini}. It consists of an array of 700 SNAILs, with the external magnetic flux alternating in polarity between adjacent unit cells. The device is closely related to those reported in \cite{Ranadive, Levochkina, ranadive2025twpa_isolator} and its schematic is illustrated in Fig.~\ref{fig:1}(d). The red inset shows neighboring SNAIL elements. 
Such configuration aims to reduce the overall 3WM nonlinearity and facilitate 4WM.
We exploit the flux tunability of the Kerr nonlinearity, together with independent single-pump gain calibration, to realize single-mode squeezing through dual-pump degenerate 4WM. In contrast to TWPAs based on dispersion engineering, the absence of gaps in the spectrum has the advantage of reduced gain ripples, and the frequency of the pumps and squeezed mode is not restricted by the design of the device \cite{Ranadive}.
Our results demonstrate single-mode squeezing at two distinct flux biases, highlighting the importance of balanced pump contributions and showing that stronger squeezing does not necessarily require higher dual-pump gain. These findings motivate further exploration of this architecture for quantum-state engineering \cite{Petrovnin} and quantum-enhanced sensing \cite{Wang2026}.

\section{Feasibility of Dual-Pump Degenerate Four-Wave Mixing} 
\label{sec:sectionII}

\rev{The present device employs reversed-Kerr phase matching \cite{Ranadive}, enabled by the negative quartic coefficient $g_4$ at $\Phi_{\mathrm{ext}}/\Phi_0=0.5$, as shown in Fig.~\ref{fig:1}(e). Within the leading-order model (Appendix~\ref{sec:A}), $g_4$ determines the 4WM coupling and contributes to self- and cross-phase modulation (SPM and XPM). The device also remains operational at other flux biases \cite{levochkina_phd_2025, Levochkina2}. In particular, at $\Phi_\mathrm{ext}/\Phi_0 = 0.0$ applied flux, it exhibits a narrow, approximately bell-shaped single-pump gain profile.}

\rev{Below, we investigate whether the present device architecture can support single-mode squeezing through dual-pump degenerate 4WM. To this end, we first examine the phase-matching conditions and then perform numerical simulations to determine whether the two pumps can provide comparable gain contributions at the signal frequency.}

\subsection{Phase-Matching Condition}

\begin{figure*}[!t]
    \centering
    \includegraphics[width=0.87\textwidth]{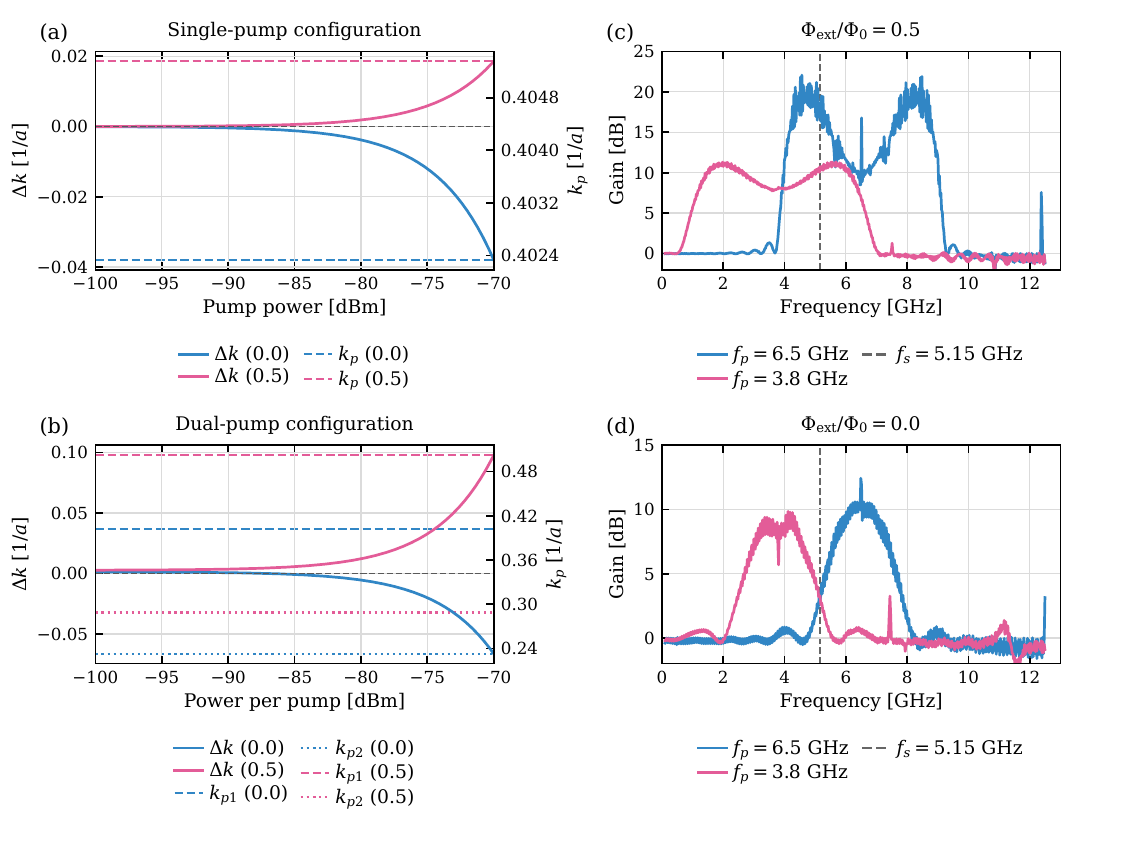}
    \caption{
    \textbf{Feasibility investigation of dual-pump degenerate 4WM.}
    (a) Calculated phase mismatch as a function of pump power
    for single-pump operation at $f_{\mathrm{p}} = 6.5$~GHz,
    with signal and idler symmetrically detuned from the pump
    by $200$~MHz.
    (b) Corresponding results for dual-pump operation with
    pump frequencies of $6.5$ and $3.8$~GHz and degenerate
    signal and idler at $5.15$~GHz for
    $\Phi_{\mathrm{ext}}/\Phi_0 = 0.0$ (blue) and $0.5$ (pink) flux biases.
    Solid lines represent the total phase mismatch, while
    dashed and dotted lines indicate the pump wavevectors
    on the right axes.
    For dual-pump operation, both pump powers are swept
    simultaneously and kept equal.
    (c) Numerically simulated TWPA gain obtained using
    JosephsonCircuits.jl package at
    $\Phi_{\mathrm{ext}}/\Phi_0 = 0.5$ and     
    (d) $\Phi_{\mathrm{ext}}/\Phi_0 = 0.0$ with
    $P_1 = -79$~dBm ($0.5~\mu$A) and $P_2 = -76$~dBm ($0.7~\mu$A).
    The vertical dashed line indicates the signal frequency
    $f_{\mathrm{s}} = 5.15$~GHz considered for dual-pump operation.
    }
    \label{fig:feasibility}
\end{figure*}

To evaluate dual-pump phase matching, we extend the model of Refs.~\cite{Ranadive,Ranadive_PhD} to two pump tones. The resulting expressions are given in Appendix~\ref{sec:A}, Eq.~\eqref{eq:a17}. We compute the phase mismatch for both single and dual-pump 4WM configurations at two representative flux bias points $\Phi_{\mathrm{ext}} / \Phi_0 = 0$ and $0.5$, \rev{corresponding to positive and negative bare quartic coefficients $g_4$, respectively}.

The transmission line is modeled as a discrete array of unit cells of length $a=8.75~\mathrm{\mu m}$ \cite{Ranadive,Ranadive_PhD}, relating the dimensionless propagation constant $k$ to the physical wavevector $k_{\mathrm{phys}}=k/a$. The phase mismatch $\Delta k$ in Fig.~\ref{fig:feasibility}(a,b) is expressed in units of $a^{-1}$ and compared with the pump wavevectors (dashed lines), obtained from the linear dispersion relation (Appendix~\ref{sec:A}, Eq.~\eqref{eq:a7}).

Figure~\ref{fig:feasibility}(a) shows the single-pump 4WM phase mismatch versus pump power, with signal and idler symmetrically detuned by 200~MHz \cite{Esposito}. At $\Phi_{\mathrm{ext}}/\Phi_0=0.5$, the mismatch remains small below approximately $-82$~dBm, reaching $\Delta k\simeq 0.02\,a^{-1}$ near saturation ($\sim-70$~dBm). At $\Phi_{\mathrm{ext}}/\Phi_0=0.0$, phase matching is less favorable. $\Delta k$ approaches zero only below $\sim-88$~dBm and reaches approximately $-0.04\,a^{-1}$ near saturation, indicating stronger phase-matching degradation with increasing pump power.

\begin{figure*}[t]
    \centering
    \includegraphics[width=0.95\textwidth]{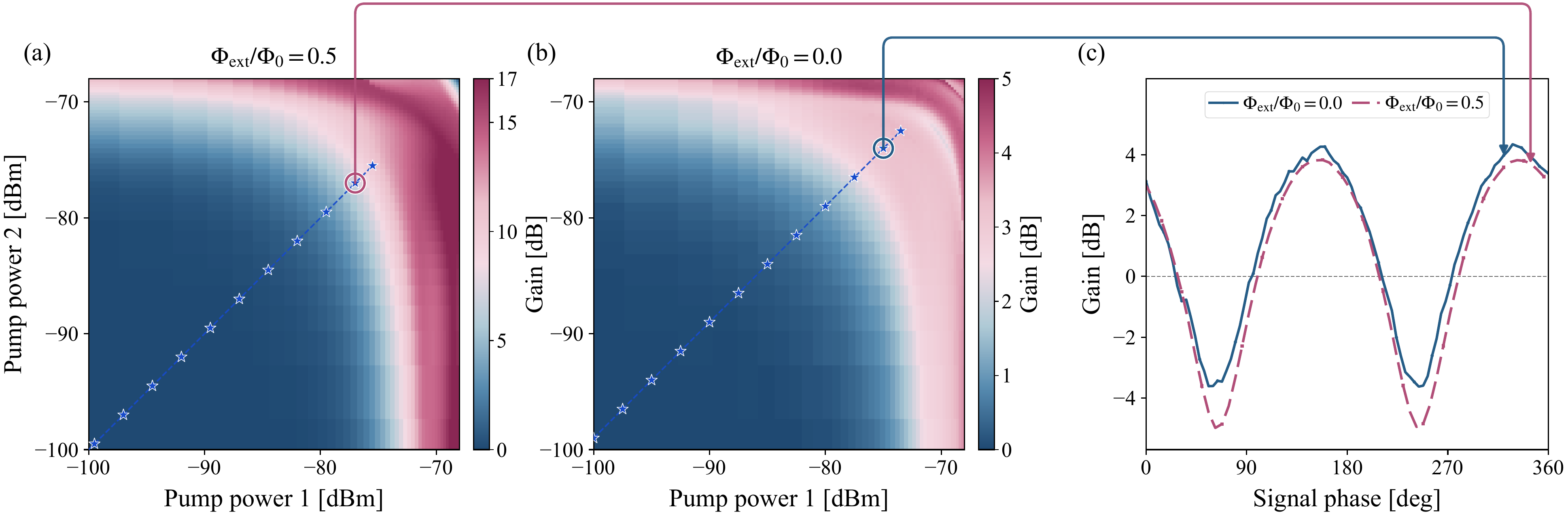}
    \caption{\textbf{Measured dual-pump gain a function of pump powers} at
    (a) $\Phi_\mathrm{ext}/\Phi_0=0.5$ and
    (b) $\Phi_\mathrm{ext}/\Phi_0=0.0$ flux.
    (c) Phase-dependent gain at $\Phi_\mathrm{ext}/\Phi_0=0.0$
    (blue solid curve) and $0.5$ (pink dashed curve) with pump powers set to $P_1=P_2=-75~\mathrm{dBm}$ for the blue curve, and $P_1=P_2=-77~\mathrm{dBm}$ for the pink curve. The signal power is $P_\mathrm{s}=-110~\mathrm{dBm}$, signal frequency $f_\mathrm{s}=5.15~\mathrm{GHz}$.}
    \label{fig:5}
\end{figure*}

Figure~\ref{fig:feasibility}(b) shows the dual-pump configuration, obtained by sweeping both pump powers simultaneously while keeping them equal. The pump frequencies are chosen as $f_{\mathrm{p}_1} = 6.5$~GHz and $f_{\mathrm{p}_2} = 3.8$~GHz, with signal and idler overlapping at $f_\mathrm{s} = 5.15$~GHz, in order to realize experimentally relevant dual-pump 4WM configurations.
At $\Phi_{\mathrm{ext}}/\Phi_0 = 0.0$, $\Delta k$ reaches approximately $-0.06\,a^{-1}$ near $-70$ $\text{dBm}$, while it is negligible at low pump powers. Notably, in the low-power regime (up to $\sim -86$ $\text{dBm}$), phase matching is slightly improved compared to the single-pump case, where $\Delta k$ remains nonzero over the same range. At $\Phi_{\mathrm{ext}}/\Phi_0=0.5$, the mismatch remains below the single-pump value up to $\sim-77~\mathrm{dBm}$, then increases more rapidly, reaching $\Delta k\simeq0.09\,a^{-1}$ near saturation.

This analysis shows that dual-pump driving preserves phase matching at low pump powers, while at higher powers the phase mismatch increases more rapidly, particularly at $\Phi_{\mathrm{ext}}/\Phi_0 = 0.5$, narrowing the optimal operating window without precluding degenerate 4WM.

\subsection{Flux-Dependent Gain in the Single-Pump Regime}

Efficient dual-pump degenerate 4WM is favored by comparable single-pump gains, which depend on the flux-dependent nonlinear response, pump frequencies, and amplitudes \cite{Brein}. To examine their relative strengths, we simulate the individual single-pump gain responses at different flux biases. For this purpose, we employ numerical simulations based on the \text{JosephsonCircuits.jl} package \cite{JosepshonCircuitjl}, which has been shown to accurately reproduce the behavior of TWPA devices \cite{Peng}.
For the simulations, we choose pump frequencies $f_{p_1}=3.8,\mathrm{GHz}$ and $f_{p_2}=6.5,\mathrm{GHz}$ to compare the corresponding gain profiles at the two flux biases. The input signal and pump tones are implemented as current sources driving the nonlinear circuit, as in Ref.~\cite{Levochkina2}. This choice allows a direct mapping between the drive amplitude and an equivalent microwave power in a $50~\Omega$ environment. 

Figures~\ref{fig:feasibility}(c) and (d) show the simulated gain profiles at $\Phi_{\mathrm{ext}}/\Phi_0=0.5$ and $0.0$, respectively, for $P_1=-79$~dBm ($0.5~\mu$A) and $P_2=-76$~dBm ($0.7~\mu$A). At $\Phi_{\mathrm{ext}}/\Phi_0=0.5$ flux, the profiles are broad and M-shaped, and the higher-frequency pump produces substantially greater single-pump gain at the degenerate signal frequency $f_s=5.15$~GHz. At $\Phi_{\mathrm{ext}}/\Phi_0=0.0$ flux, the profiles are narrower and bell-shaped, with weaker but more comparable single-pump gains at $f_s$ for the same applied powers. \rev{These examples illustrate that in such device architecture the relative single-pump gains depend on both flux bias and pump settings. The difference observed at $\Phi_{\mathrm{ext}}/\Phi_0=0.5$ flux can therefore be addressed by adjusting the pump powers, with independent single-pump calibrations providing a practical guide for selecting dual-pump operating conditions.}

\section{Experimental Characterization of Single Mode Squeezing} 
\label{sec:sectionIII}

To experimentally investigate single-mode squeezing, the device was mounted at the $20$ $\text{mK}$ stage of a BlueFors dilution refrigerator, using the experimental setup described in Appendix~\ref{sec:B}. Signal generation and acquisition were performed using a Presto instrument~\cite{Presto} in its different operating modes. All measurements were carried out in a dual-pump configuration with $f_{\text{p}_1} = 3.8~\mathrm{GHz}$ and $f_{\text{p}_2} = 6.5~\mathrm{GHz}$, corresponding to a degenerate signal frequency $f_{\text{s}} = 5.15~\mathrm{GHz}$, at two flux points $\Phi_\mathrm{ext}/\Phi_0 = 0.0$ and $0.5$. 

\subsection{Dual-Pump Gain in Degenerate 4WM Regime}

The dual-pump gain in the degenerate 4WM regime was first characterized using Presto continuous-wave (Lockin) operation mode \cite{Presto}. A weak probe tone at the signal frequency $f_\text{s}$ was applied with a power $P_\text{s} = -110$~dBm, and the gain was measured using a pump ``ON -- OFF'' method \cite{Levochkina}. The measured gain landscapes, obtained by sweeping both pump powers, reveal distinct regions of parametric amplification at $\Phi_\mathrm{ext}/\Phi_0 = 0.5$ and $0.0$ (Fig.~\ref{fig:5}(a,b)), with maximum gains of approximately $17$ and $5$~dB, respectively.
\rev{However, the observed amplification does not necessarily reflect balanced dual-pump 4WM, since the individual pump contributions may differ substantially. The blue star trace in Figs.~\ref{fig:5}(a) and (b) indicate pump-power combinations expected to yield comparable single-pump gains at $f_\text{s}$, as estimated from Appendix~\ref{sec:F}, Fig.~\ref{fig:app_gain}. Owing to calibration uncertainties, these points represent approximate operating conditions, and other combinations may provide similar gain balance.}

To further investigate the phase-sensitive response, we explored different operating points along the star-marked trajectories, observing phase-dependent amplification. Figure~\ref{fig:5}(c) presents representative measurements at one operating point per flux bias, chosen for appreciable gain ($P_1=P_2=-75$~dBm at $\Phi_\mathrm{ext}/\Phi_0=0.0$ and $P_1=P_2=-77$~dBm at $\Phi_\mathrm{ext}/\Phi_0=0.5$). 
Unlike phase-preserving amplification, the dual-pump TWPA exhibits a periodic dependence on the probe phase, alternating between amplification and deamplification. The phase-sensitive extinction ratio (PSER), defined as the difference between the maximum and minimum gain in decibels, reaches $\sim 8$~dB at both flux biases. Although modest compared with previous reports~\cite{Brein}, these results demonstrate phase-sensitive amplification, providing a basis for the squeezing measurements presented below.

\subsection{Measurement of Single-Mode Squeezing}

To characterize single-mode squeezing, we employ Presto operating in pulsed mode with 0 Hz intermediate frequency \cite{Presto}. This allows both signal generation and acquisition at $f_s$ via direct digital synthesis without external mixers. The signal port of the TWPA is connected to a heavily attenuated input line (Fig.~\ref{fig:ap2}) terminated with a $50~\Omega$ load, which suppresses room-temperature noise and thermalizes the incident field toward the cryostat base temperature. At $f_s$, the residual thermal occupation is assumed to be negligible, so the input field is treated, to a good approximation, as a quantum vacuum state.

\begin{figure}[t!]
    \hspace*{0.00\textwidth} 
    \includegraphics[width=0.5\textwidth]{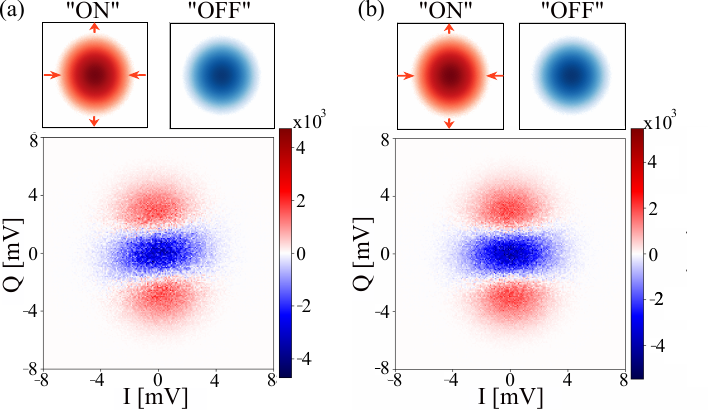} 
    \caption{
    \textbf{Examples of quadrature histograms} 
    Histograms were constructed from $4 \times 10^{7}$ $IQ$ samples for (a) $\Phi_{\mathrm{ext}}/\Phi_0 = 0.5$ and $(P_1,P_2) \approx (-77,-77)$~dBm and (b) $\Phi_{\mathrm{ext}}/\Phi_0 = 0.0$ and $(P_1,P_2) \approx (-75,-75)$~dBm. The upper panels show the elliptical pump ``ON'' distributions (red) and approximately circular pump ``OFF'' reference distributions (blue). The lower panels show their differences.}
    \label{fig:7}
\end{figure}

Two pump tones at frequencies $f_{\text{p}1} = 3.8~\mathrm{GHz}$ and $f{\text{p}_2} = 6.5~\mathrm{GHz}$ are generated as square pulses. \rev{The sequence alternates between $10~\mu\mathrm{s}$ pump ``ON'' and ``OFF'' intervals, allowing comparison with the unpumped reference state.} For both flux points, the resulting output signal is digitized using the Presto ADC at sampling rates up to $1~\mathrm{GS/s}$, with an effective bandwidth set by the applied filtering, yielding the in-phase ($I$) and quadrature ($Q$) components of the field. \rev{To exclude switching transients, only the central $8~\mu\mathrm{s}$ of each pulse is analyzed. Two-dimensional histograms of the digitized $I$ and $Q$ voltages are constructed from $4\times10^7$ $IQ$ samples each.} \rev{Figure~\ref{fig:7} shows representative distributions at the operating points used for the phase-sensitive gain measurements in Fig.~\ref{fig:5}(c).} \rev{Without pumping, the distributions are approximately circular and centered at the origin, consistent with phase-independent reference noise} (Fig. \ref{fig:7}(a) and (b), top right). \rev{With both pumps applied, they become elliptical, revealing phase-dependent quadrature fluctuations} (Fig.~\ref{fig:7}(a) and (b), top left). The principal axes of the ellipse correspond to directions of \rev{minimum and maximum quadrature variance}. \rev{The lower panels show pump ``ON -- OFF''  difference histograms, illustrating the redistribution of quadrature fluctuations. To quantify squeezing relative to the vacuum level, Johnson–Nyquist thermometry (Appendix~\ref{sec:C}) calibrates the system gain and input-referred added noise, enabling reconstruction of the quadrature variances (Appendix~\ref{sec:D}).}

\begin{figure}[!t]
    \hspace*{0.00\textwidth}
    \includegraphics[width=0.42\textwidth]{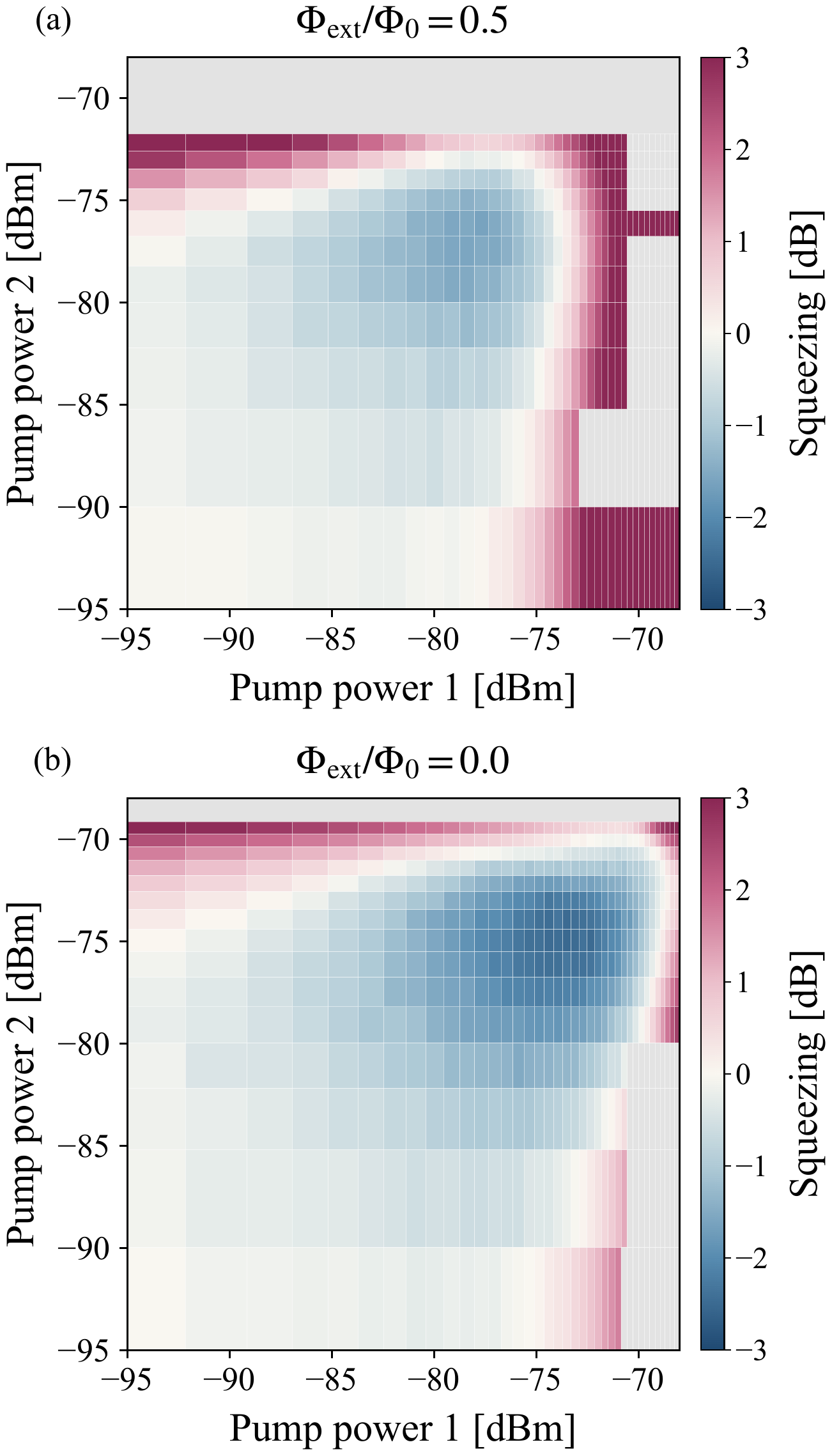}
    \caption{
    \textbf{Measurement results of single-mode squeezing.}
    Colormaps of the squeezing level, defined as the minimum calibrated quadrature variance relative to vacuum (in dB), as a function of pump powers $P_1$ and $P_2$ at (a) $\Phi_{\mathrm{ext}}/\Phi_0 = 0.5$ and (b) $\Phi_{\mathrm{ext}}/\Phi_0 = 0.0$. Grey regions correspond to pump-power combinations not covered by the measurements.}
    \label{fig:8}
\end{figure}

\rev{Following a similar approach to Ref.~\cite{Brein}, we mapped the calibrated squeezing as a function of both pump powers, sweeping one while holding the other fixed. The resulting maps at $\Phi_{\mathrm{ext}}/\Phi_0=0.5$ and $0.0$ are shown in Figs.~\ref{fig:8}(a) and (b), respectively.} The \rev{strongest measured squeezing} is $-1.62$ $\mathrm{dB}$ and $-2.48$  $\mathrm{dB}$ at the corresponding flux biases. \rev{The squeezing maps exhibit different pump-power dependences at the two flux biases, with squeezing persisting to higher pump powers at $\Phi_{\mathrm{ext}}/\Phi_0=0.0$. Further characterization of the squeezed and anti-squeezed quadratures and inferred Gaussian-state purity is provided in Appendix~\ref{sec:D}, while the relation between squeezing, total gain, and individual pump contributions is examined in the following section.}

\begin{figure*}[t!]
    \centering
    \includegraphics[
        width=\textwidth,
        height=\textheight,
        keepaspectratio
    ]{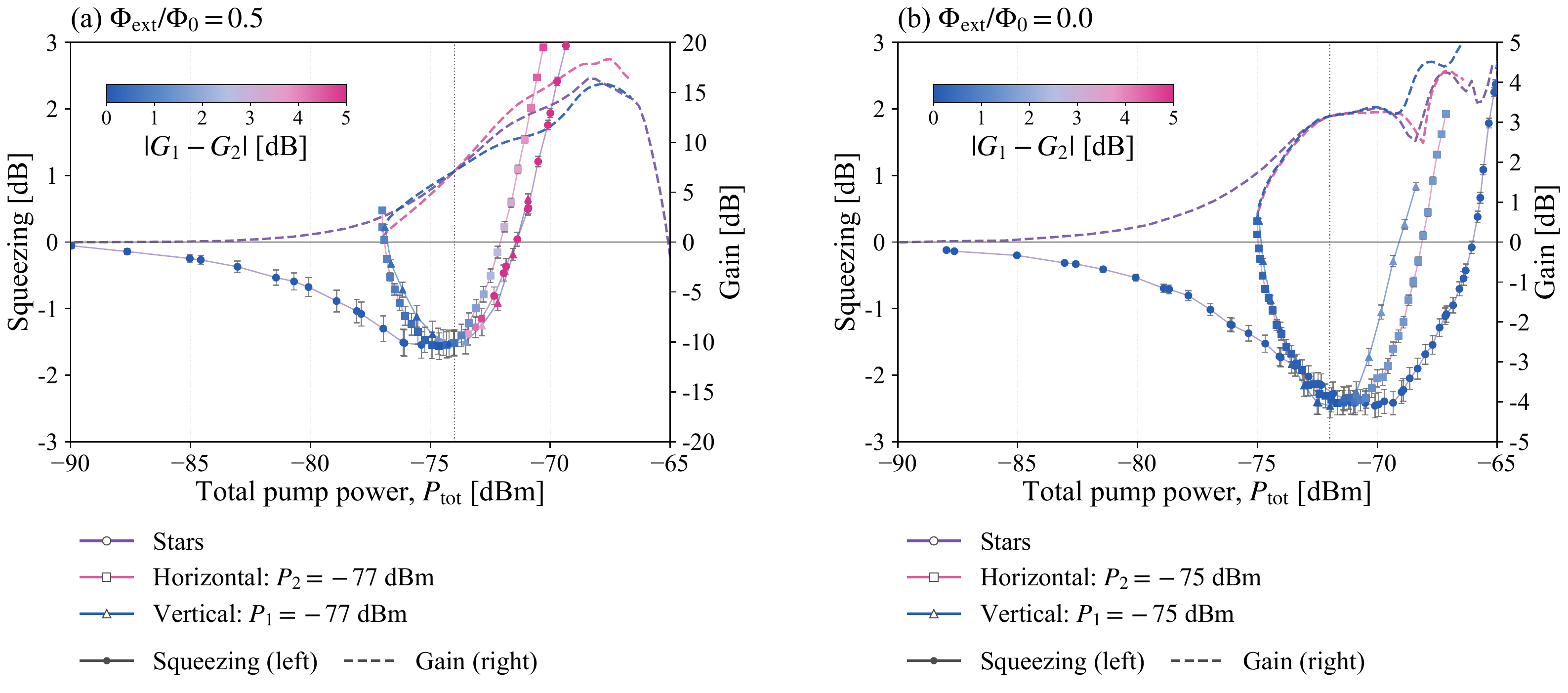}
\caption{\textbf{Squeezing analysis.} \rev{Measured squeezing (markers, left axis) and dual-pump gain (dashed curves, right axis) as functions of total pump power for $\Phi_{\mathrm{ext}}/\Phi_0=0.5$ (a) and $0.0$ (b).
    Each panel shows three pump-power trajectories: approximately balanced single-pump gains (stars from Fig. \ref{fig:5}) and horizontal and vertical cuts through the region of strongest squeezing.
    Marker colors indicate the gain imbalance $|G_1(P_1)-G_2(P_2)|$, obtained from independent single-pump calibrations at $f_s$, with a common color scale from $0$ to $5~\mathrm{dB}$.
    The horizontal dotted line denotes the vacuum noise reference.
    }}
    \label{fig:squeezing_analysis}
\end{figure*}

\vspace{-6pt}
\section{Results and Discussion}

\rev{
Figure~\ref{fig:squeezing_analysis} compares measured squeezing and dual-pump gain along three pump-power trajectories, namely, approximately balanced single-pump gains (the star trajectory in
Fig.~\ref{fig:5}(a,b)), and horizontal and vertical
cuts through the region of strongest squeezing.
Marker colors indicate the gain imbalance
$|G_1(P_1)-G_2(P_2)|$, obtained from independent
single-pump calibrations at $f_s$
(Fig.~\ref{fig:app_gain}).}
\rev{At $\Phi_{\mathrm{ext}}/\Phi_0=0.5$ (Fig.~\ref{fig:squeezing_analysis}(a)), the strongest squeezing is $-1.62\pm0.20~\mathrm{dB}$ at approximately $5.5~\mathrm{dB}$ gain, whereas at
$\Phi_{\mathrm{ext}}/\Phi_0=0.0$ (Fig.~\ref{fig:squeezing_analysis}(b)), it reaches
$-2.48\pm0.18~\mathrm{dB}$ at approximately
$3~\mathrm{dB}$ gain. The uncertainties are dominated by the system-gain calibration $G_{\mathrm{sys}}$. The stronger squeezing at $\Phi_{\mathrm{ext}}/\Phi_0=0.0$ flux may partly reflect more favorable phase matching. The calculated mismatch in Fig.~\ref{fig:feasibility}(b) is smaller in magnitude there over part of the lower-power range. However, the present measurements do not isolate this contribution from other flux-dependent effects.}

\rev{Along the approximately balanced trajectory, squeezing develops gradually from lower total pump powers. The horizontal and vertical cuts begin at higher total powers with no squeezing, but approach similar minimum variances as the other pump power is increased. At $\Phi_{\mathrm{ext}}/\Phi_0=0.5$, the three trajectories largely overlap near the optimum and during the subsequent deterioration, while at $\Phi_{\mathrm{ext}}/\Phi_0=0.0$, their high-power responses separate more clearly. At both biases, squeezing deteriorates despite increasing
dual-pump gain, even along the approximately balanced trajectory, indicating that the calibrated single-pump gain imbalance alone cannot explain the upturn.}

\rev{Several mechanisms may contribute to the degradation beyond the optimal pump power. Higher-order nonlinearities may alter the effective Kerr response and phase matching \cite{Sivak2019Kerr}, while residual three-wave mixing cannot be excluded experimentally despite the vanishing $g_3$ nonlinearity predicted at both symmetry points \cite{Levochkina}. Internal loss and unwanted mixing processes may further limit squeezing \cite{Esposito_perspectives}, including pump-dependent spurious spectral tones recently reported in a reversed-Kerr SNAIL TWPA \cite{levochkina_spurious_2026}. Appendix~\ref{sec:E} illustrates how internal pump imbalance can affect squeezing and phase matching within an idealized model, without quantitatively describing the experiment.}

\indent

\section{Conclusion}
\label{sec:conclusions}
In conclusion, we have demonstrated single-mode squeezing at two flux biases using dual-pump degenerate four-wave mixing in a reversed-Kerr traveling-wave parametric amplifier.
By avoiding engineered spectral gaps, this architecture offers greater freedom in choosing the pump and squeezed-mode frequencies.
Guided by independent single-pump gain calibrations, we investigated pump configurations expected to provide comparable gains at $f_s$. The strongest measured squeezing was $-1.62\pm0.20$~dB at $\Phi_{\mathrm{ext}}/\Phi_0=0.5$
and $-2.48\pm0.18$~dB at $\Phi_{\mathrm{ext}}/\Phi_0=0.0$.

\rev{The stronger squeezing obtained at zero flux and lower dual-pump gain points to the importance of flux-dependent operating conditions. At both biases, the high-power deterioration persists even for approximately
balanced single-pump gains. Its origin therefore remains an open question, requiring further investigation of the nonlinear response, loss, and additional mixing processes beyond the reduced model considered here.}

Nevertheless, these results highlight the potential of this architecture as a source of squeezed microwave states for quantum information processing and quantum-enhanced measurements.


\section*{Acknowledgements} 
\label{sec:acknowledgements}
This project has received funding from the European Union under Horizon Europe 2021-2027 Framework Programme, project MiSS (Microwave Squeezing with Superconducting (meta)materials) grant agreement ID: 101135868. 
We also acknowledge the financial support of the Finnish Ministry of Education and Culture through the Quantum Doctoral Education Pilot Program (QDOC VN/3137/2024-OKM-4) and the Research Council of Finland through the Finnish Quantum Flagship project 358877 at Aalto University.
We are also grateful for support from Saab, under a research collaboration agreement with Aalto University. We thank Silent Waves for providing the Carthago TWPA sample used in this work. We acknowledge the provision of facilities and technical support by Aalto University at OtaNano - Low Temperature Laboratory.
\section*{Data Availability Statement}

The data that support the findings of this study are available from the corresponding author upon reasonable request.

\appendix
\renewcommand{\thesection}{\Alph{section}}
\renewcommand{\section}[1]{%
  \refstepcounter{section}%
  \vspace{1em} 
  \begin{center}
    \textbf{Appendix \thesection: #1}
  \end{center}
  \medskip
}


\appendix
\section{Phase Matching in the Dual-Pump Degenerate 4WM Regime}
\label{sec:A}

\begin{figure*}[t!]
    \centering
    \hspace*{-0.06\textwidth}
    \includegraphics[
        width=\textwidth,
        height=0.45\textheight,
        keepaspectratio
    ]{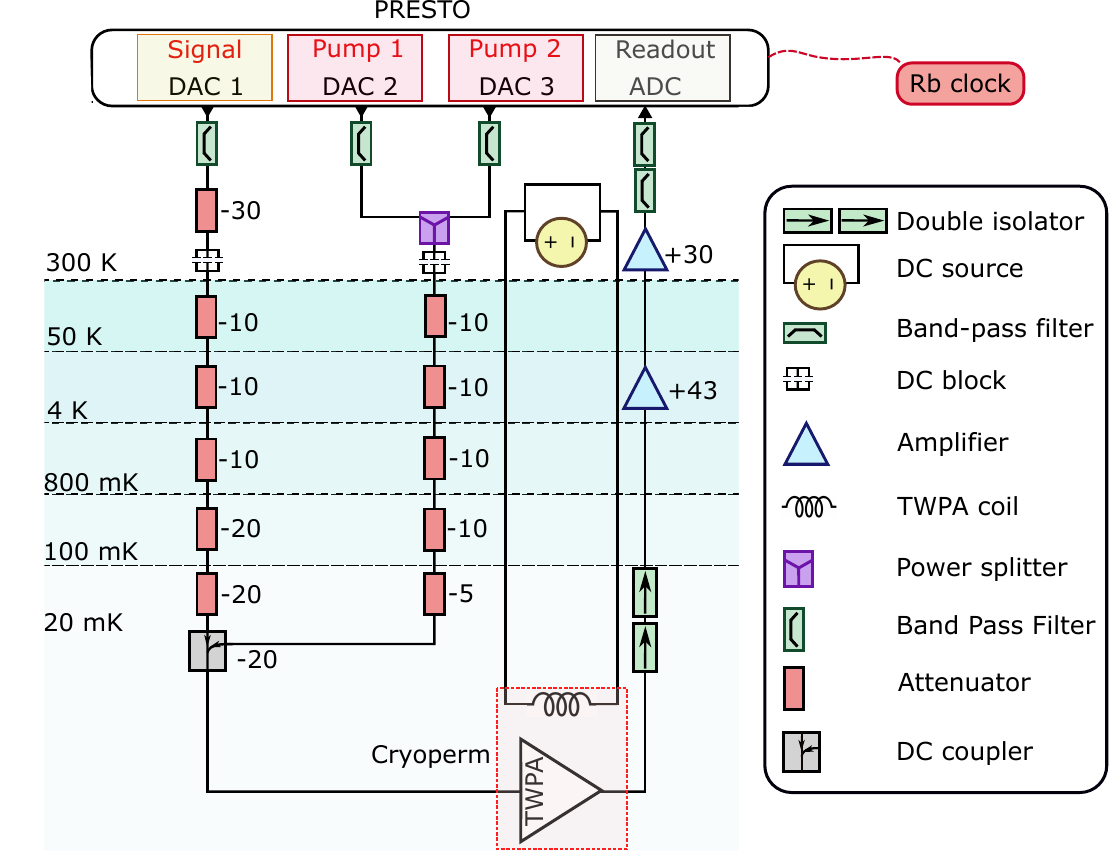}
    \caption{\textbf{Schematic of the dual-pump experimental
    setup for characterizing single-mode squeezing.}}
    \label{fig:ap2}
\end{figure*}

We describe phase matching in a uniform SNAIL-based
transmission line, assuming undepleted classical pumps, slowly varying envelopes, and negligible effective 3WM \cite{Ranadive_PhD,Ranadive}.

For a SNAIL containing three identical large junctions ($I_C$) and one smaller junction ($rI_C$) as shown in Fig. \ref{fig:1}(d), the current--phase relation is
\begin{equation}
 \frac{I_L(\varphi)}{I_C}
 =r\sin\varphi+
 \sin\left(\frac{\varphi-\varphi_{\mathrm{ext}}}{3}\right),
 \label{eq:a0}
\end{equation}
where $\varphi_{\mathrm{ext}}=2\pi\Phi_{\mathrm{ext}}/\Phi_0$
is the normalized external flux.
Expanding Eq.~\eqref{eq:a0} about a stable equilibrium $\varphi^*$,
with $I_L(\varphi^*)=0$, gives
\begin{equation}
 \frac{I_L(\varphi^*+\varphi)}{I_C}
 =\tilde\alpha\varphi-\tilde\beta\varphi^2
 -\tilde\gamma\varphi^3+O(\varphi^4),
 \label{eq:a1}
\end{equation}
where
\begin{align}
 \tilde\alpha
 &=r\cos\varphi^*
 +\frac13\cos\left(
 \frac{\varphi^*-\varphi_{\mathrm{ext}}}{3}\right),
 \label{eq:a2}\\
 \tilde\beta
 &=\frac12\left[
 r\sin\varphi^*
 +\frac19\sin\left(
 \frac{\varphi^*-\varphi_{\mathrm{ext}}}{3}\right)
 \right],
 \label{eq:a3}\\
 \tilde\gamma
 &=\frac16\left[
 r\cos\varphi^*
 +\frac1{27}\cos\left(
 \frac{\varphi^*-\varphi_{\mathrm{ext}}}{3}\right)
 \right].
 \label{eq:a4}
\end{align}
We define
$\gamma=\tilde\gamma/\tilde\alpha$,
$L=\Phi_0/(2\pi I_C\tilde\alpha)$, and
$E_C=e^2/(2C_g)$, where $C_g$ is the ground capacitance
per cell \cite{Levochkina}. In our convention, the four-wave mixing flux-tunable nonlinear coefficient is
\begin{equation}
 \hbar g_4=\frac{\tilde\gamma}{2\tilde\alpha}E_C.
 \label{eq:a5}
\end{equation}

Alternating flux polarity suppresses coherent 3WM
in the ideal long-wavelength limit \cite{Ranadive}.
We neglect residual 3WM (assuming that $g_3 = 0$) and cascaded corrections.

\rev{
The propagation coordinate $x=z/a_{\mathrm{cell}}$ counts
unit cells, where $z$ is physical distance and
$a_{\mathrm{cell}}$ is the cell length.
The device extends from $x=0$ to $x=N$.
Wavevectors, nonlinear phase shifts, and parametric
couplings are therefore expressed per cell.
}

The dimensionless node-phase envelopes are defined by
\begin{equation}
 \phi(x,t)=\frac12\sum_m
 \left[A_m(x)e^{i(k_mx-\omega_mt)}+\mathrm{c.c.}\right].
 \label{eq:phase_envelope_definition}
\end{equation}
\rev{
Thus, the wave equation can be written as
}
\begin{equation}
 \rev{
 \partial_x^2\phi-\omega_0^{-2}\partial_t^2\phi
 +\omega_J^{-2}\partial_x^2\partial_t^2\phi
 -\gamma\partial_x[(\partial_x\phi)^3]=0,
 }
 \label{eq:appA_nonlinear_wave}
\end{equation}
where $\omega_0=1/\sqrt{LC_g}$,
$\omega_J=1/\sqrt{LC_J}$ and
$C_J$ denotes the effective junction capacitance per cell.
The linear part gives
\begin{equation}
 k(\omega)=\frac{\omega}
 {\omega_0\sqrt{1-\omega^2/\omega_J^2}}.
 \label{eq:a7}
\end{equation}

We define nonlinear wavevector corrections by
$A_m(x)\propto e^{i\alpha_m x}$, so that
$q_m=k_m+\alpha_m$ \cite{obrien2014resonant}.
For a single pump, the phase mismatch is
\begin{equation}
 \Delta k=q_s+q_i-2q_p
 =k_s+k_i-2k_p+\alpha_s+\alpha_i-2\alpha_p.
 \label{eq:a6}
\end{equation}
Retaining resonant cubic terms gives
\begin{align}
 \alpha_p
 &=\frac{3\gamma\omega_0^2}{8\omega_p^2}
 k_p^5|A_p|^2,
 \label{eq:a9}\\
 \alpha_m
 &=\frac{3\gamma\omega_0^2}{4\omega_m^2}
 k_m^3k_p^2|A_p|^2,\qquad m=s,i.
 \label{eq:a10}
\end{align}

For two distinct pumps, the degenerate signal satisfies
\begin{equation}
 \omega_s=\frac{\omega_{p_1}+\omega_{p_2}}2,
 \qquad
 \Delta k_{\mathrm{lin}}=2k_s-k_{p_1}-k_{p_2}.
 \label{eq:a12}
\end{equation}
Writing $U_j=|A_{p_j}|^2$, and neglecting phase shifts
produced by the weak signal, pump self- and cross-phase
modulation give \cite{white2015traveling}
\begin{equation}
 \alpha_{p_j}
 =\frac{3\gamma\omega_0^2k_{p_j}^3}{8\omega_{p_j}^2}
 \left(k_{p_j}^2U_j+2k_{p_\ell}^2U_\ell\right),
 \quad (j,\ell)=(1,2),(2,1),
 \label{eq:a14}
\end{equation}
while the signal phase shift is
\begin{equation}
 \alpha_s
 =\frac{3\gamma\omega_0^2k_s^3}{4\omega_s^2}
 \left(k_{p_1}^2U_1+k_{p_2}^2U_2\right).
 \label{eq:a15}
\end{equation}
Eq.~\eqref{eq:a6} can be rewritten taking into account Eq.~\eqref{eq:a12} as
\begin{equation}
 \Delta k=\Delta k_{\mathrm{lin}}
 +2\alpha_s-\alpha_{p_1}-\alpha_{p_2} = \Delta k_{\mathrm{lin}} + \Delta k_{\mathrm{NL}}.
 \label{eq:a17}
\end{equation}
Equivalently, defining
$F_m=3\gamma\omega_0^2k_m^3/(8\omega_m^2)$,
\begin{align}
 \Delta k_{\mathrm{NL}}&=c_1U_1+c_2U_2,
 \label{eq:a16}\\
 c_1&=k_{p_1}^2(4F_s-F_{p_1}-2F_{p_2}),\\
 c_2&=k_{p_2}^2(4F_s-2F_{p_1}-F_{p_2}).
\end{align}

\rev{
The cubic nonlinear term in
Eq.~\eqref{eq:appA_nonlinear_wave},
which originates from the quartic Josephson
interaction, also generates the degenerate
four-wave-mixing coupling between the two pumps
and the signal, consistent with the squeezing
Hamiltonian in Eq.~(17) of
Ref.~\cite{Grimsmo}.
Retaining the resonant contribution
$A_{p_1}A_{p_2}A_s^*$ gives
}
\begin{equation}
 \rev{
 \frac{dA_s}{dx}
 =i\alpha_s A_s
 +i\mathcal C A_{p_1}(x)A_{p_2}(x)A_s^*
 e^{-i\Delta k_{\mathrm{lin}}x},
 }
 \label{eq:signal_envelope_coupling}
\end{equation}
\rev{
where, with the envelope normalization of
Eq.~\eqref{eq:phase_envelope_definition},
}
\begin{equation}
 \rev{
 \mathcal C
 =\frac{3\gamma\omega_0^2}{4\omega_s^2}
 k_s^2k_{p_1}k_{p_2}(k_{p_1}+k_{p_2}-k_s).
 }
 \label{eq:coupling_coefficient}
\end{equation}

\rev{
The coupling prefactor follows from the six
resonant permutations of $A_{p_1}A_{p_2}A_s^*$ in the cubic nonlinear term, using the linear dispersion relation. Derivatives of the slowly varying envelopes
in the nonlinear source are neglected.
}

\rev{
For undepleted pumps,
$A_{p_j}(x)=\bar A_{p_j}e^{i\alpha_{p_j}x}$,
with constant $U_j=|\bar A_{p_j}|^2$.
Removing the signal nonlinear phase through
$\mathcal A_s=A_se^{-i\alpha_s x}$ yields
}
\begin{equation}
\rev{
\frac{d\mathcal A_s}{dx}
=i\chi e^{-i\Delta kx}\mathcal A_s^*,
\qquad
\chi=\mathcal C\bar A_{p_1}\bar A_{p_2}.
}
\label{eq:degenerate_coupling}
\end{equation}
\rev{
where $\chi$
is the complex parametric coupling per cell,
distinct from the bare quartic coefficient
$g_4$. This form is consistent with the
standard four-wave-mixing squeezing
description~\cite{Grimsmo}.
}

\section{Experimental setup}
\label{sec:B}

The experimental setup, designed to generate and detect microwave signals required to probe the dual-pump 4WM process in the TWPA, is depicted in Fig.~\ref{fig:ap2}.

To control and measure the TWPA device, we employ a Presto arbitrary waveform generator and analyzer \cite{Presto}. Presto operates in the second Nyquist band, enabling direct digital synthesis of signals up to 10 $\text{GHz}$ with an instantaneous bandwidth of 500 $\text{MHz}$. The signal and pump tones are generated directly by the system and routed through dedicated digital-to-analog converter (DAC) channels. Specifically, DAC~1 generates the signal at $f_s = 5.15$~GHz. This signal is passed through a band-pass filter (Mini-Circuits VBFZ-5500-S+), with a passband from 4.9 to 6.2 $\text{GHz}$ and subsequently attenuated by 30 $\text{dB}$ at room temperature. DAC~2 produces the first pump tone at $f_{p_1} = 3.8$ $\text{GHz}$, which is filtered using a band-pass filter (Mini-Circuits VBFZ-3590-S+) with a passband from 3.0 to 4.3 $\text{GHz}$. DAC~3 generates the second pump tone at $f_{p_2} = 6.5$~GHz, followed by filtering with a Mini-Circuits VHF-3500A+ band-pass filter covering 3.9 to 7.0~GHz. To ensure balanced pumping conditions, we perform a preliminary calibration of the two pump channels such that the powers at $f_{p_1}$ and $f_{p_2}$ at the cryostat input are equal. To compensate for frequency-dependent variations in the output chain, an additional 4~dB digital attenuation is applied to the lower-frequency pump channel.

The input lines are attenuated at multiple temperature stages inside the cryostat to thermalize the signals and suppress thermal noise before reaching the device, as shown in Fig.~\ref{fig:ap2}. The TWPA is enclosed in a Cryoperm-10 magnetic shield to minimize flux noise and ensure stable operation. 

At the output, the signal is first amplified by a cryogenic HEMT amplifier at 4~K (Low Noise Factory LNF-LNC4\_8A), followed by additional low-noise amplification stages at room temperature with net gain of 30~dB. The amplified signal is then passed through a band-pass filter (Mini-Circuits VBFZ-5500-S+), with a passband from 4.9 to 6.2~GHz, and an additional high-pass filter (Mini-Circuits VHF-4600+) to suppress low-frequency components, before being digitized by the Presto ADC via direct RF sampling at $f_s = 5.15$~GHz for further processing.

An external Keysight 33250A arbitrary waveform generator (AWG) is used to apply the flux bias to the TWPA, enabling in-situ tuning of the device nonlinearity and phase-matching conditions.

\rev{
For the chosen experimental frequencies,
additional four-wave-mixing channels may arise at
$f_s\pm|f_{p_2}-f_{p_1}|$,
corresponding to $2.45$ and $7.85\,\mathrm{GHz}$.
While the experimental filtering suppresses
their contribution to the detected signal,
it does not necessarily eliminate their
coupling within the device.
These channels are not explicitly included
in the reduced theoretical model and may
become relevant at higher pump powers.
}

\section{System Gain and Noise Calibration}
\label{sec:C}

The calibration follows the standard framework of quantum-limited linear amplifiers \cite{Clerk_10}, implemented via a Johnson–Nyquist noise thermometry measurement. The output noise power is recorded as a function of the physical temperature $T$ of the 20 $\text{mK}$ stage, assuming a fully thermalized $50~\Omega$ environment. The calibration is performed at the signal frequency $f_s = 5.15~\mathrm{GHz}$.

The Johnson–Nyquist noise power at the device input is
\begin{equation}
P_{\mathrm{in}}(T) = k_B \, \mathrm{BW} \, T,
\end{equation}
where $\mathrm{BW} = 1~\mathrm{MHz}$ is the measurement bandwidth.

After amplification by the full measurement chain, the measured output power is
\begin{equation}
P_{\mathrm{out}}(T) = k_B \, \mathrm{BW} \, G_{\mathrm{sys}} \left(T + T_{\mathrm{add}}\right),
\end{equation}
where $G_{\mathrm{sys}}$ is the total system gain and $T_{\mathrm{add}}$ is the system-added noise temperature referred to the input.

In practice, the measured output power is also expressed in an equivalent temperature scale via
\begin{equation}
T_{\mathrm{eq}} \equiv \frac{P_{\mathrm{out}}}{k_B \, \mathrm{BW}},
\end{equation}
so that
\begin{equation}
T_{\mathrm{eq}} = G_{\mathrm{sys}} \left(T + T_{\mathrm{add}}\right).
\end{equation}

\begin{figure}[!t]
    \hspace*{-0.05\textwidth}
    \includegraphics[width=0.43\textwidth]{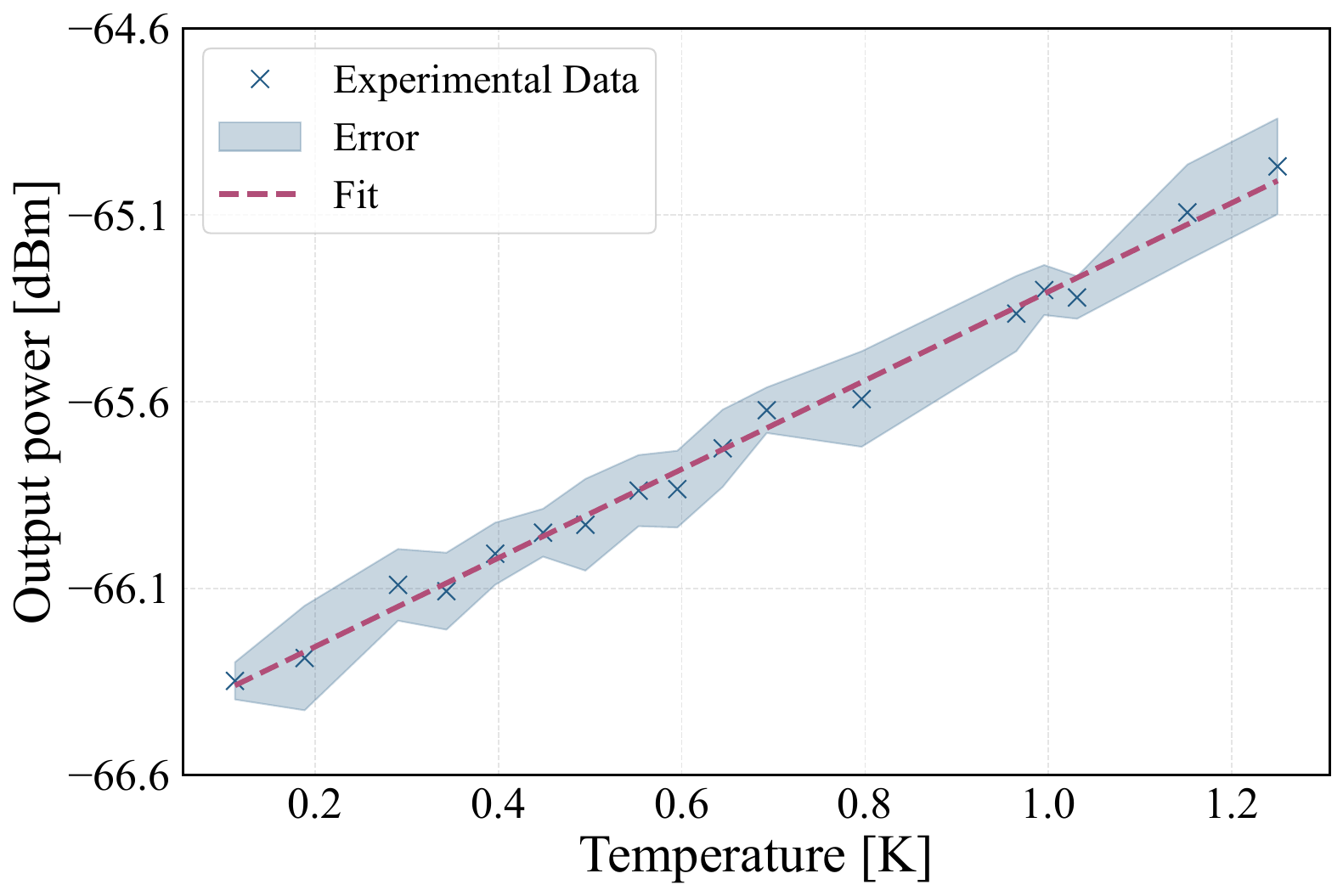} 
    \caption{\textbf{Measured output power as a function of the physical temperature $T$ of the input load.} A linear fit is used to extract the total system gain $G_{\mathrm{sys}}$ and the system-added noise temperature $T_{\mathrm{add}}$.}
    \label{fig:ap3}
\end{figure}

A linear fit of $T_{\mathrm{eq}}$ versus $T$ therefore yields both $G_{\mathrm{sys}}$ and $T_{\mathrm{add}}$ (Fig. \ref{fig:ap3}).

From this calibration we obtain
\begin{equation}
G_{\mathrm{sys}} = (67.31 \pm 0.26)~\mathrm{dB}, \quad T_{\mathrm{add}} = (2.97 \pm 0.17)~\mathrm{K}.
\end{equation}
The uncertainties are derived from the standard error of the linear fit and propagated to the dB scale using standard error propagation for the logarithmic transformation. For the quadrature calibration in the squeezing measurements, all measured voltage variances are normalized using the previously extracted $G_{\mathrm{sys}}$ and $T_{\mathrm{add}}$.

\section{Single-Pump Gain Calibration}
\label{sec:F}

\begin{figure}[h!]
    \hspace*{-0.04\textwidth}
    \includegraphics[width=0.43\textwidth]{
        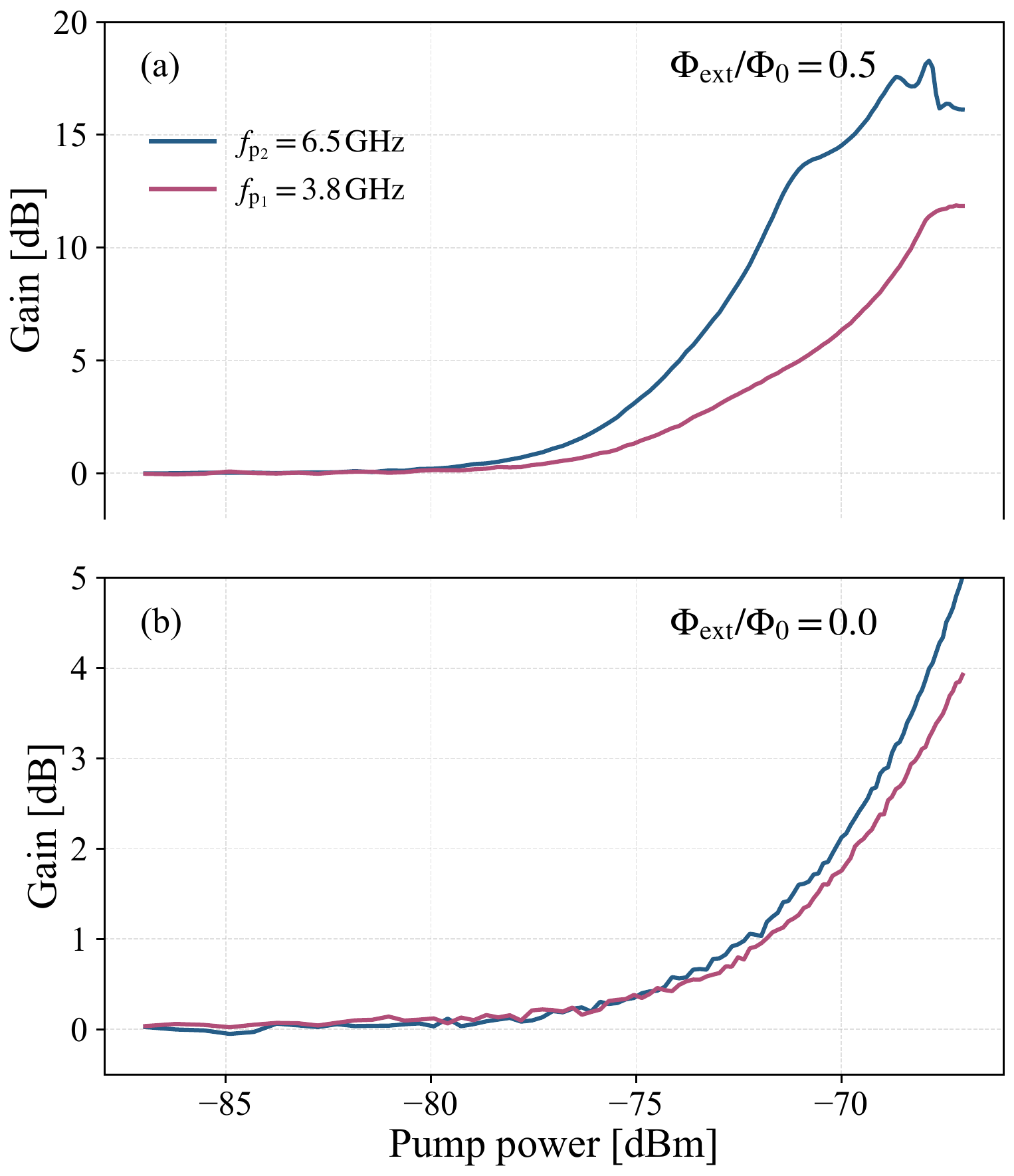
    }
    \caption{\textbf{Measured individual-pump gain as a function
    of pump power} at
    (a) $\Phi_{\mathrm{ext}}/\Phi_0=0.5$ and
    (b) $\Phi_{\mathrm{ext}}/\Phi_0=0.0$.}
    \label{fig:app_gain}
\end{figure}

Figures~\ref{fig:app_gain} (a) and (b) show the experimentally measured gain at frequency $f_s$ for the two flux bias points $\Phi_{\mathrm{ext}}/\Phi_0 = 0.5$ and $\Phi_{\mathrm{ext}}/\Phi_0 = 0.0$, respectively. In each case, the gain is measured independently for the two pump tones $f_{p_1} = 3.8~\mathrm{GHz}$ (pink curves) and $f_{p_2} = 6.5~\mathrm{GHz}$ (blue curves). The signal frequency is kept fixed at $f_s = 5.15~\mathrm{GHz}$, while signal power $P_s$ is set to $-110$ $\text{dBm}$. These calibrations allow us to compare the effective amplification provided by each pump under identical operating conditions, forming the basis for the imbalance metric introduced in the main text.

\section{Quadrature Reconstruction and Squeezing Calibration}
\label{sec:D}

We measure the output field using a Presto ADC, which
provides time traces of the in-phase $I(t)$ and quadrature $Q(t)$ components. For each pump configuration (ON and OFF), the data are segmented into repeated acquisition records and digitally demodulated at the signal frequency $f_s$.

Before the measurements, the Presto full-scale amplitude was calibrated against the microwave power measured with a spectrum analyzer, allowing the recorded quadratures to be expressed on a calibrated power scale. All noise-power calculations are performed in linear units, using
\begin{equation}
 P[\mathrm{W}]=10^{-3}10^{P[\mathrm{dBm}]/10}.
\end{equation}

We define the complex envelope and rotated quadrature as
\begin{equation}
 X(t)=I(t)+iQ(t),
 \qquad
 X_\theta(t)=I(t)\cos\theta+Q(t)\sin\theta.
\end{equation}
For each pump configuration, we evaluate
\begin{equation}
 \operatorname{Var}(X_\theta)
 =\langle X_\theta^2\rangle-\langle X_\theta\rangle^2.
\end{equation}
The minimum and maximum over $\theta$ identify the
squeezed and anti-squeezed quadratures, $X_-$ and $X_+$.

For comparison with quantum fluctuations, we use
the canonical quadratures
\begin{equation}
 \hat X=\frac{\hat a+\hat a^\dagger}{\sqrt{2}},
 \qquad
 \hat P=\frac{\hat a-\hat a^\dagger}{i\sqrt{2}},
\end{equation}
whose vacuum variances are
\begin{equation}
 \operatorname{Var}(X_{\mathrm{vac}})
 =\operatorname{Var}(P_{\mathrm{vac}})=\frac12.
\end{equation}

The reconstructed variances are obtained using the independently calibrated detection-chain power gain ($G_{\mathrm{sys}}$) (Appendix~\ref{sec:C}). The measured noise power is divided by this gain in linear units and converted to canonical quadrature units using the vacuum noise scale ($hf_s\mathrm{BW}/2$), with the normalization appropriate to the measured quadrature components.

For a phase-insensitive detection chain, the measured variance, referred to the detection-chain input, contains both the field variance and the added detection noise:
\begin{equation}
 \operatorname{Var}(V_i)
 =\operatorname{Var}(X_i)+n_{\mathrm{add}},
 \qquad
 n_{\mathrm{add}}=\frac{k_BT_{\mathrm{add}}}{hf_s},
\end{equation}
where $T_{\mathrm{add}}$ is the calibrated equivalent
input-referred added-noise temperature.

Assuming that the pump-OFF field at the reference plane
approximates vacuum,
\begin{equation}
 \operatorname{Var}(V_{\mathrm{off}})
 =\frac12+n_{\mathrm{add}}.
\end{equation}
Eliminating the common added noise gives
\begin{equation}
 \operatorname{Var}(X_i)
 =
 \operatorname{Var}(V_{i,\mathrm{on}})
 -\operatorname{Var}(V_{i,\mathrm{off}})
 +\frac12,
 \qquad i=\pm.
 \label{eq:quad_reconstruct}
\end{equation}
This subtraction assumes unchanged detection-chain gain and added noise between ON and OFF acquisitions. It does not remove fluctuations generated within the pumped TWPA, and no correction based on an assumed internal device loss is applied.

The squeezing and anti-squeezing levels are
\begin{equation}
 S_{\pm,\mathrm{dB}}=10\log_{10}S_\pm,
 \qquad
 S_\pm=
 \frac{\operatorname{Var}(X_\pm)}
 {\operatorname{Var}(X_{\mathrm{vac}})}
 =2\operatorname{Var}(X_\pm).
 \label{eq:quad_squeezing}
\end{equation}

For a single-mode Gaussian state, the purity $\mu=\operatorname{Tr}(\rho^2)$ is determined by the covariance matrix $V$ in the $(X,P)$ basis
\cite{Paris2003PurityGaussian}.
In its principal-axis basis,
\begin{equation}
 V=
 \begin{pmatrix}
  \operatorname{Var}(X_-) & 0\\
  0 & \operatorname{Var}(X_+)
 \end{pmatrix},
\end{equation}
and therefore
\begin{equation}
 \mu=\frac{1}{2\sqrt{\det V}}
 =\frac{1}{\sqrt{S_-S_+}}.
 \label{eq:quad_purity}
\end{equation}
This expression gives the inferred purity under the assumption that the selected measurement mode is Gaussian \cite{Weedbrook2012}.

\begin{figure}[!t]
    \centering
    \includegraphics[width=0.43\textwidth]{
        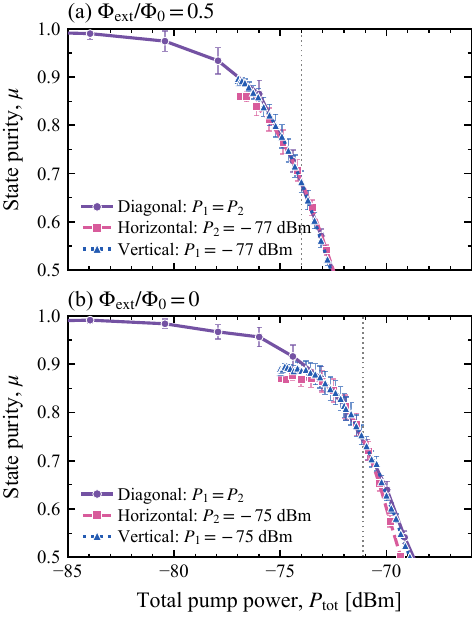
    }
    \caption{
    \textbf{Pump-power dependence of the inferred
    Gaussian-state purity.}
    Purity calculated from the reconstructed squeezed
    and anti-squeezed variances along nominally balanced (star trajectory), horizontal (fixed $P_2$), and vertical (fixed $P_1$) cuts through the pump-power maps for (a) $\Phi_{\mathrm{ext}}/\Phi_0=0.5$ and (b) $\Phi_{\mathrm{ext}}/\Phi_0=0.0$ flux. The horizontal and vertical cuts
    intersect at $P_1=P_2=-77$~dBm in (a) and
    $P_1=P_2=-75$~dBm in (b).
    Vertical dotted lines indicate the corresponding
    total applied pump powers.
    }
    \label{fig:9}
\end{figure}

\rev{
Figure~\ref{fig:9}(a,b) presents the inferred Gaussian-state purity, calculated using Eq.~\eqref{eq:quad_purity}, at flux biases $\Phi_{\mathrm{ext}}/\Phi_0=0.5$ and $0.0$, respectively, along the pump-power trajectories shown in Fig.~\ref{fig:squeezing_analysis}. Combining the reconstructed squeezed and corresponding anti-squeezed quadrature variances
yields purities of $0.727\pm0.039$ and $0.668\pm0.028$ at the operating points of maximum squeezing for the respective flux biases. The purity exhibits broadly similar trends along the three trajectories at each flux bias,
generally decreasing with increasing total pump power.
For an ideal pure Gaussian state, $S_-S_+=1$ and $\mu=1$, whereas the
experimentally reconstructed variances are not constrained to satisfy this relation. The observed deviations from unity indicate mixedness within the Gaussian-state description, without uniquely identifying its physical origin.
}

\section{Pump Imbalance and Distributed Loss}
\label{sec:E}

\rev{To gain further insight into the observed squeezing
behavior, we examine pump imbalance using the
leading-order model of Appendix~\ref{sec:A}
and consider the influence of distributed loss
and associated vacuum fluctuations.
The resulting analytical expressions help clarify the squeezing
behavior under these simplified conditions.}

\rev{
Using Eqs.~\eqref{eq:coupling_coefficient}
and~\eqref{eq:degenerate_coupling},
we consider a fixed total pump intensity
$U=U_1+U_2$ and define the imbalance parameter
$d=(U_1-U_2)/U$. The coupling strength and
phase mismatch can then be written as
}
\begin{align}
 \rev{|\chi|}
 &\rev{=\frac{|\mathcal C|U}{2}\sqrt{1-d^2},}
 \\
 \rev{\Delta k(d)}
 &\rev{=\Delta k_{\mathrm{lin}}
 +\frac{U}{2}\left[(c_1+c_2)+(c_1-c_2)d\right],}
 \label{eq:appF_imbalance}
\end{align}

\rev{
For the lossless calculation, we retain a single
normalized degenerate signal mode, with
$[\hat a,\hat a^\dagger]=1$, and assume constant
coefficients. Its evolution follows the coupling
convention of Eq.~\eqref{eq:degenerate_coupling},
consistent with the standard description of
degenerate parametric amplification
\cite{Grimsmo,CollettGardiner1984}:
}
\begin{equation}
 \rev{
 \frac{d\hat a}{dx}
 =i\chi e^{-i\Delta kx}\hat a^\dagger.
 }
 \label{eq:appF_mode_evolution}
\end{equation}
\rev{
Introducing $\hat b=e^{i\Delta kx/2}\hat a$ gives
}
\begin{equation}
 \rev{
 \frac{d\hat b}{dx}
 =i\frac{\Delta k}{2}\hat b+i\chi\hat b^\dagger.
 }
 \label{eq:appF_bogoliubov_evolution}
\end{equation}
\rev{
This equation is generated by
$d\hat b/dx=i[\hat{\mathcal K},\hat b]$, with
}
\begin{equation}
 \rev{
 \hat{\mathcal K}
 =-\frac{\Delta k}{2}\hat b^\dagger\hat b
 -\frac12\left(
 \chi\hat b^{\dagger 2}+\chi^*\hat b^2
 \right).
 }
 \label{eq:appF_generator}
\end{equation}
\rev{
Here $\hat{\mathcal K}$ is a Hermitian generator
of spatial evolution, expressed per cell rather
than as an energy. The factor $1/2$ follows from
$[\hat b^{\dagger 2},\hat b]=-2\hat b^\dagger$.
The resulting quadratic interaction describes
single-mode squeezing through four-wave mixing
\cite{Grimsmo}.
}

\begin{figure}[t!]
 \centering
 \vspace{-0.2pt}
 \includegraphics[width=0.45\textwidth]{
 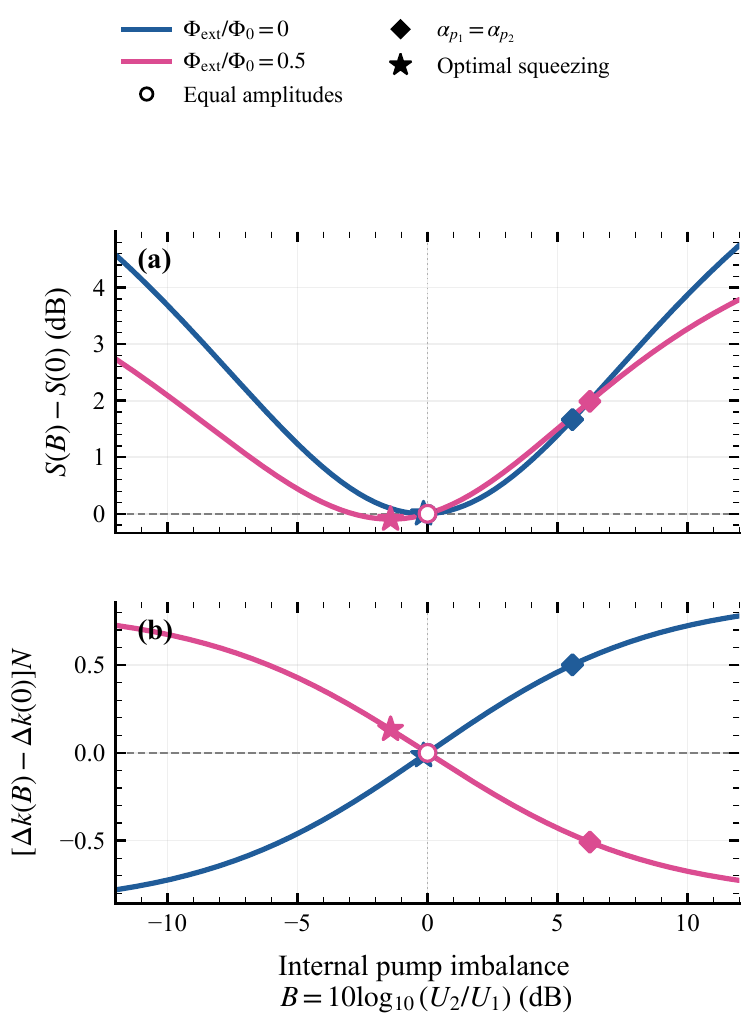
 }
 \caption{\rev{
 \textbf{Pump imbalance in the lossless model.}
 Calculations at $\Phi_{\mathrm{ext}}/\Phi_0=0$
 (blue) and $0.5$ (orange), for fixed $U_1+U_2$,
 versus $10\log_{10}(U_2/U_1)$.
 (a) Change in squeezing,
 $\delta S(B)=10\log_{10}[S_-(B)/S_-(0)]$;
 negative values indicate improved squeezing.
 (b) Change in accumulated mismatch,
 $[\Delta k(B)-\Delta k_{\mathrm{bal}}]N$.
 Circles mark equal pump intensities, diamonds
 equal nonlinear pump shifts
 ($\alpha_{p_1}=\alpha_{p_2}$), and stars
 minimum quadrature variance.
 }}
 \label{fig:imbalance}
\end{figure}

\rev{
For constant coefficients, the solution takes
the Bogoliubov form
$\hat b(N)=u\hat b(0)+v\hat b^\dagger(0)$,
with \cite{Grimsmo,CollettGardiner1984}
}
\begin{align}
 \rev{u}
 &\rev{=\cosh(gN)+i\frac{\Delta k}{2g}\sinh(gN),}
 \\
 \rev{v}
 &\rev{=i\frac{\chi}{g}\sinh(gN),}
 \\
 \rev{g^2}
 &\rev{=|\chi|^2-\left(\frac{\Delta k}{2}\right)^2.}
 \label{eq:appF_bogoliubov_coefficients}
\end{align}
\rev{
The coefficients satisfy $|u|^2-|v|^2=1$,
preserving the bosonic commutation relation.
}

\rev{
For vacuum input, the output quadrature
$\hat X_\theta=
(e^{-i\theta}\hat b+e^{i\theta}\hat b^\dagger)/\sqrt{2}$
has variance
}
\begin{equation}
 \rev{
 2\operatorname{Var}(\hat X_\theta)
 =|u|^2+|v|^2+
 2\operatorname{Re}(uv e^{-2i\theta}).
 }
\end{equation}
\rev{
Minimizing over the quadrature angle $\theta$
gives the vacuum-normalized squeezed variance
\cite{Grimsmo,CollettGardiner1984}
}
\begin{equation}
 \rev{
 S_-=(|u|-|v|)^2=e^{-2r},
 }
 \label{eq:appF_ideal_variance}
\end{equation}
\rev{where}
\begin{equation}
 \rev{
 r=\operatorname{arcsinh}|v|
 =\operatorname{arcsinh}
 \left|\frac{\chi}{g}\sinh(gN)\right|.
 }
 \label{eq:appF_squeezing_parameter}
\end{equation}
\rev{
Thus $S_-=1$ for vacuum, and at perfect
phase matching, $r=|\chi|N$.
}

\rev{
At fixed total internal pump intensity,
equal pump intensities maximize the coupling
magnitude $|\chi|$. However, when $c_1\neq c_2$,
a finite imbalance can reduce the magnitude
of the total phase mismatch $|\Delta k|$.
The squeezing optimum therefore reflects
a competition between coupling strength and
phase matching. The model retains nonlinear
phase corrections arising from pump self- and
cross-phase modulation. Equalizing the nonlinear
pump shifts,
$\alpha_{p_1}=\alpha_{p_2}$,
does not generally maximize squeezing, since
the total mismatch also contains linear
dispersion and the nonlinear signal phase shift.
}

\rev{
Figure~\ref{fig:imbalance} illustrates the
physical significance of this competition
within the reduced lossless model. The
calculation is intended to distinguish
the conditions of equal internal pump
intensities, equal nonlinear pump shifts,
and minimum quadrature variance, which
need not coincide. At
$\Phi_{\mathrm{ext}}/\Phi_0=0$, the minimum
variance occurs near equal pump intensities,
whereas at $0.5$ a small imbalance slightly improves
the calculated squeezing. This illustrates how
a reduction in phase mismatch can compensate
for a weaker parametric coupling. The figure thus illustrates the influence of pump imbalance on squeezing without aiming to quantitatively reproduce the experimentally observed optimum.
}

\rev{
For comparison, we consider a simplified analytical
benchmark for distributed loss under perfect phase
matching, constant coupling, and independent vacuum
reservoirs \cite{CollettGardiner1984,Houde2019}.
The power attenuation coefficient
$\alpha_{\mathrm{loss}}\geq0$ is defined by
}
\begin{equation}
 \rev{
 P(x)=P(0)e^{-\alpha_{\mathrm{loss}}x}
 }
 \label{eq:appF_power_attenuation}
\end{equation}
\rev{
in the absence of pumping, corresponding to a field
amplitude decay rate of $\alpha_{\mathrm{loss}}/2$
per cell. For physical distance $z=a_{\mathrm{cell}}x$,
the attenuation coefficient becomes
$\alpha_z=\alpha_{\mathrm{loss}}/a_{\mathrm{cell}}$,
in inverse-length units.
}

\rev{
At perfect phase matching, a constant phase rotation
defines a signal operator $\hat c$ with lossless
evolution $d\hat c/dx=\rho\hat c^\dagger$,
where $\rho=|\chi|$. Including attenuation and the
associated vacuum fluctuations gives the spatial
Heisenberg--Langevin equation
\cite{CollettGardiner1984,Houde2019}
}
\begin{equation}
 \rev{
 \frac{d\hat c}{dx}
 =\rho\hat c^\dagger
 -\frac{\alpha_{\mathrm{loss}}}{2}\hat c
 +\sqrt{\alpha_{\mathrm{loss}}}\,\hat f(x).
 }
 \label{eq:appF_langevin}
\end{equation}
\rev{
The zero-mean vacuum noise satisfies
}
\begin{equation}
 \rev{
 [\hat f(x),\hat f^\dagger(x')]
 =\langle\hat f(x)\hat f^\dagger(x')\rangle
 =\delta(x-x'),
 }
\end{equation}
\rev{
with
$\langle\hat f^\dagger(x)\hat f(x')\rangle
=\langle\hat f(x)\hat f(x')\rangle=0$.
These reservoir fluctuations preserve the bosonic
commutation relation in the presence of attenuation.
}

\rev{
For the squeezed quadrature
$\hat X_-=(\hat c-\hat c^\dagger)/(i\sqrt{2})$,
the evolution becomes
}
\begin{equation}
 \rev{
 \frac{d\hat X_-}{dx}
 =-\left(\rho+\frac{\alpha_{\mathrm{loss}}}{2}\right)
 \hat X_-+\sqrt{\alpha_{\mathrm{loss}}}\,\hat X_f,
 }
\end{equation}
\rev{
where
$\hat X_f=(\hat f-\hat f^\dagger)/(i\sqrt{2})$
has symmetrized correlation
$\langle\{\hat X_f(x),\hat X_f(x')\}\rangle/2
=\delta(x-x')/2$.
The vacuum-normalized variance
$S_-=2\operatorname{Var}(\hat X_-)$
therefore obeys
}
\begin{equation}
 \rev{
 \frac{dS_-}{dx}
 =-(2\rho+\alpha_{\mathrm{loss}})S_-
 +\alpha_{\mathrm{loss}},
 \qquad S_-(0)=1.
 }
 \label{eq:loss_variance}
\end{equation}
\rev{
Here, parametric squeezing and attenuation reduce
the quadrature variance, while vacuum fluctuations
introduced through the loss channel contribute
the additive term $\alpha_{\mathrm{loss}}$.
This is consistent with the standard vacuum-loss
channel description \cite{CollettGardiner1984}.
}

\rev{
For constant coefficients, integration yields
}
\begin{equation}
 \rev{
 \begin{aligned}
 S_-(N)={}&e^{-(2\rho+\alpha_{\mathrm{loss}})N}\\
 &+\frac{\alpha_{\mathrm{loss}}}
 {2\rho+\alpha_{\mathrm{loss}}}
 \left[1-e^{-(2\rho+\alpha_{\mathrm{loss}})N}\right].
 \end{aligned}
 }
 \label{eq:loss_squeezed}
\end{equation}
\rev{
This expression preserves the vacuum variance
under pure attenuation ($\rho=0$) and recovers
$S_-=e^{-2\rho N}$ without loss, with the
zero-coupling, zero-loss case understood by continuity.
For fixed positive coefficients, increasing $N$
approaches the limiting variance
$\alpha_{\mathrm{loss}}/(2\rho+\alpha_{\mathrm{loss}})$.
At fixed attenuation and propagation length,
increasing $\rho$ improves squeezing monotonically,
so this simplified model does not produce a
high-power squeezing upturn.
}

\rev{
The analytical treatment of distributed loss
provides a complementary perspective on the
limitations of quadrature squeezing. Under the
assumptions of perfect phase matching, constant
coupling, and vacuum reservoirs, attenuation
introduces a limiting variance but does not
produce a squeezing upturn as the coupling
increases. This simplified analysis is separate
from the pump-imbalance calculations and does
not establish the origin of the experimentally
observed high-power deterioration. The experimental
squeezing is reconstructed using the calibrated
detection chain (Appendix~\ref{sec:D}), without
applying an assumed correction for internal
TWPA attenuation.
}

\bibliography{references}

\end{document}